\pdfoutput=1
\documentclass[11pt]{article}
\usepackage{ifpdf}
\usepackage{hyperref}
\hypersetup{pdfstartview=FitH, colorlinks=true, urlcolor=blue, citecolor=blue, linkcolor=blue}
\usepackage[margin=1in]{geometry}
\usepackage{url}
\usepackage{graphicx}
\usepackage[latin1]{inputenc}
\usepackage[T1]{fontenc}
\usepackage{times}
\usepackage{amsfonts}
\usepackage{authblk}
\usepackage{booktabs}
\usepackage{mathtools}
\usepackage{amsmath}
\usepackage{amsthm} 
\usepackage{xspace}
\usepackage{enumitem} 
\usepackage{multicol}
\usepackage{tikz}
\usepackage{float}
\usepackage{array}
\usepackage{footnote}

\theoremstyle{definition}
\newtheorem{definition}{Definition}

\newtheorem{remark}{Remark}

\newtheorem{innercustomthm}{Theorem}
\newenvironment{customthm}[1]
  {\renewcommand\theinnercustomthm{#1}\innercustomthm}
  {\endinnercustomthm}

\newcommand{\mathcmd}[1]{\ensuremath{#1}\xspace}

\renewcommand{\gets}{\mathcmd{\leftarrow}}
\newcommand{\getsr}{\mathcmd{\xleftarrow{\$}}}
\newcommand{\give}{\mathcmd{\rightarrow}}
\newcommand{\dent} {\mathcmd{\hspace*{9pt}}}
\newcommand{\rw}[1]{{\textbf{#1}}}
\newcommand{\ret} {\mathcmd{\rw{return}\hspace*{3pt}}}

\newcommand{\iif} {\mathcmd{\rw{if}\hspace*{3pt}}}

\newcommand{\then} {\mathcmd{\rw{then}\hspace*{3pt}}}
\newcommand{\eelse} {\mathcmd{\rw{else}\hspace*{3pt}}}
\newcommand{\doo} {\mathcmd{\rw{do}\hspace*{3pt}}}

\newcommand{\foor} {\mathcmd{\rw{for}\hspace*{3pt}}}

\newcommand{\flag} {\mathcmd{\mathtt{flag}}}
\newcommand{\role} {\mathcmd{\mathtt{role}}}

\newcommand{\getscup}{\mathcmd{\xleftarrow{\cup}}}

\newcommand{\Adv}{\mathcmd{\mathbf{Adv}}}
\newcommand{\Exp}{\mathcmd{\mathbf{Exp}}}
\newcommand{\advA} {\mathcmd{\mathcal{A}}}

\newcommand{\atka} {\mathcmd{\mathsf{Enroll \sdash Valid Key \sdash Illegitimate Entity}}}
\newcommand{\atkb} {\mathcmd{\mathsf{Enroll \sdash Invalid Key \sdash Illegitimate Entity}}}
\newcommand{\atkc} {\mathcmd{\mathsf{Enroll \sdash Invalid Key \sdash Legitimate Entity}}}
\newcommand{\atkd} {\mathcmd{\mathsf{Update \sdash Collision}}}
\newcommand{\atke} {\mathcmd{\mathsf{Revoke \sdash Collision}}}

\newcommand{\setup} {\mathcmd{\mathsf{Setup}}} 
\newcommand{\register} {\mathcmd{\mathsf{Enroll}}}

\newcommand{\verify} {\mathcmd{\mathsf{Verify}}}
\newcommand{\update} {\mathcmd{\mathsf{Update}}}
\newcommand{\revoke} {\mathcmd{\mathsf{Revoke}}}

\newcommand{\addBL} {\mathcmd{\mathsf{AddBL}}}

\newcommand{\Opt} {\mathcmd{\mathsf{Opt}}}

\newcommand{\accgen} {\mathcmd{\mathsf{AccGen}}}
\newcommand{\accadd} {\mathcmd{\mathsf{AccAdd}}}
\newcommand{\accwitadd} {\mathcmd{\mathsf{AccWitAdd}}}
\newcommand{\accver} {\mathcmd{\mathsf{AccVer}}}
\newcommand{\upmsg} {\mathcmd{\mathsf{updmsg}}}
\newcommand{\accdel} {\mathcmd{\mathsf{AccDel}}}
\newcommand{\accmemwitupondel} {\mathcmd{\mathsf{MemWitUpdateDel}}}

\newcommand{\pk} {\mathcmd{\mathsf{pk}}}
 
\newcommand{\regpkspace} {\boldsymbol{\mathcmd{\mathsf{PK}}}}
\newcommand{\pkspace} {\boldsymbol{\mathcmd{\mathcal{PK}}}}
\newcommand{\regidspace} {\boldsymbol{\mathcmd{\mathsf{ID}}}}
\newcommand{\ledger} {\mathcmd{\mathcal{L}}}
\newcommand{\sk} {\mathcmd{\mathsf{sk}}}
\newcommand{\id} {\mathcmd{\mathsf{ID}}}

\newcommand{\thr} {\mathcmd{\mathsf{t}}}
\newcommand{\nrTI} {\mathcmd{\mathsf{T}}}
\newcommand{\blockchain} {\mathcmd{\mathsf{BL}}}

\newcommand{\templist} {\mathcmd{\mathsf{TL}}} 
\newcommand{\keygen} {\mathcmd{\mathsf{KG}}}
\newcommand{\pkveri} {\mathcmd{\mathsf{PIV}}}
\newcommand{\msg} {\mathcmd{\mathrm{m}}}

\newcommand{\sign} {\mathcmd{\mathsf{Sign}}}
\newcommand{\veri} {\mathcmd{\mathsf{Veri}}} 
\newcommand{\acc} {\mathcmd{\mathrm{a}}}
\newcommand{\PI} {\mathcmd{\mathsf{PI}}}
\newcommand{\SI} {\mathcmd{\mathsf{SI}}} 
\newcommand{\TI} {\mathcmd{\mathsf{TI}}} 
\newcommand{\RP} {\mathcmd{\mathsf{RP}}}
\newcommand{\EP} {\mathcmd{\mathsf{EP}}} 
\newcommand{\hash} {\mathcmd{\mathsf{H}}} 

\newcommand{\sdash} {\mathcmd{\mbox{-}}}
\newcommand{\eufcma} {\mathcmd{\mathsf{EUF\sdash CMA}}}

\newcommand{\prepare} {\mathcmd{\mathsf{prepare}}}
\newcommand{\preprepare} {\mathcmd{\mathsf{pre\sdash prepare}}}
\newcommand{\commit} {\mathcmd{\mathsf{commit}}}

\newcommand{\scheme} {\mathcmd{\mathrm{DBPKI}}} 
\newcommand{\tx} {\mathcmd{\mathrm{tx}}}

\newcommand{\ur} {\mathcmd{\mathrm{R}}}
\newcommand{\ui} {\mathcmd{\mathrm{I}}}
\newcommand{\uo} {\mathcmd{\mathrm{O}}}

\definecolor{lime}{HTML}{A6CE39}
\DeclareRobustCommand{\orcidicon}{
	\begin{tikzpicture}
	\draw[lime, fill=lime] (0,0) 
	circle [radius=0.16] 
	node[white] {{\fontfamily{qag}\selectfont \tiny ID}};
	\draw[white, fill=white] (-0.0625,0.095) 
	circle [radius=0.007];
	\end{tikzpicture}
	\hspace{-2mm}
}

\foreach \x in {A, ..., Z}{
	\expandafter\xdef\csname orcid\x\endcsname{\noexpand\href{https://orcid.org/\csname orcidauthor\x\endcsname}{\noexpand\orcidicon}}
}

\title{A Decentralized Dynamic PKI based on Blockchain}

\author{Mohsen Toorani\orcidA{} \hspace{4cm} Christian Gehrmann\\
\texttt{mohsen.toorani@eit.lth.se} \hspace{1cm} \texttt{christian.gehrmann@eit.lth.se} \\ 
\vspace{10pt}
Department of Electrical and Information Technology\\ Lund University, Sweden}
\date{}

\begin{document}
\maketitle
\let\thefootnote\relax\footnotetext{\scriptsize{Copyright  \copyright ~Authors 2020. All Rights Reserved. This is the author's version of the work. It is posted for personal use, not for redistribution. A variant will appear in SAC'21, https://doi.org/10.1145/3412841.3442038}.}

\begin{abstract}
The central role of the certificate authority (CA) in traditional public key infrastructure (PKI) makes it fragile and prone to compromises and operational failures. Maintaining CAs and revocation lists is demanding especially in loosely-connected and large systems. Log-based PKIs have been proposed as a remedy but they do not solve the problem effectively. We provide a general model and a solution for decentralized and dynamic PKI based on a blockchain and web of trust model where the traditional CA and digital certificates are removed and instead, everything is registered on the blockchain. Registration, revocation, and update of public keys are based on a consensus mechanism between a certain number of entities that are already part of the system. Any node which is part of the system can be an auditor and initiate the revocation procedure once it finds out malicious activities. Revocation lists are no longer required as any node can efficiently verify the public keys through witnesses. 
\end{abstract}

\pagestyle{plain}
\pagenumbering{arabic}
\setlength{\emergencystretch}{8em}

\section{Introduction}
\label{sec:introduction}
PKI has been used for a long time for secure channel establishment and protection of data objects. 
In the traditional PKI, the CA has a central role and is responsible for enrolling and revoking public keys. 
This makes the PKI ecosystem fragile and prone to attacks and operational failures: several CAs have been compromised, and many fraudulent certificates were issued and used for man-in-the-middle attacks \cite{KUBILAYKM19}. 
Moreover, maintaining CAs, and distribution and storage of certificate revocation lists (CRLs) is demanding, especially in loosely-connected and large systems. 

Alternatives to the CA-based PKIs include the web of trust (WoT), identity-based encryption (IBE), and certificateless public key cryptography (CL-PKC).
WoT provides a decentralized trust model where the authenticity of public keys and their owners is based on the amount of trust that other entities in WoT have in the entity in question. This removes the need for a CA but has shortcomings in providing non-repudiation, addressing revocation, and the need for a priori trust relations. 

Despite problems and limitations in the centralized trust model of the traditional PKI, it is scalable and cost-effective and provides integrity, authentication, and non-repudiation. It is then worth-investigating further improvements to the PKI model than replacing it with its alternatives. 
Several new PKI models have then been proposed, mainly based on \emph{public logs} or use of \emph{blockchain}: 

\begin{enumerate}[leftmargin=*]
\item \textbf{Log-based PKIs}: In this approach which is mainly developed for addressing the problem of misbehaving CAs, certificates are not deemed valid until they are logged on append-only public logs. 
\emph{Certificate Transparency} (CT) \cite{rfc6962} was the first log-based solution, proposed by Google for improving the accountability of CA operations and detecting mis-issued certificates. 
Anyone can audit CA activities by monitoring CT logs and notice the issuance of suspect certificates that are not in public logs. The main proposal \cite{rfc6962}, however, does not specify any mechanism for revocation and authenticity of logged certificates, but there are proposals for handling revocation \cite{Ryan14}. The idea was then extended for building a transparent PKI as in ARPKI \cite{BasinCKPSS14} 
and PoliCert \cite{SzalachowskiMP14}, and for providing transparency in messaging systems as in CIRT \cite{Ryan14}, CONIKS \cite{MelaraBBFF15}, and EthIKS \cite{Bonneau16a}. 
Such extensions typically require some extra (trusted) entities in addition to the requirement for a public log \cite{YuCR16, BasinCKPSS14, KimHPJG13}.  
Log-based PKIs have several drawbacks: They require a centralized and consistent source of information to operate securely, do not sufficiently incentivize recording or monitoring CA behaviors, and require time and manual efforts for reporting CA misbehavior. 

\item {\textbf{Blockchain-based PKIs}: 
Blockchain is an append-only database of signed transactions that can be used to eliminate central points-of-failure by providing a decentralized solution and provide certificate transparency and revocation, and reliable transaction records. From the trust model perspective, two approaches have been taken in proposals for blockchain-based PKI: 
\begin{itemize}
    \item \textbf{Hierarchical}:  
    Most proposals for blockchain-based PKI have a hierarchical trust model where some CAs will decide about certification or revocation of keys \cite{WangLCWJZ18, KUBILAYKM19, lewisonC16, yangSW18, KhieuM19, MatsumotoR16}. They are indeed log-based PKIs that use blockchain as an append-only public bulletin. Some proposals, however, involve a few more entities than only CAs and provide some relaxed centralization but still with limited functionalities since trusted central authorities are not eliminated. 
    \item \textbf{Web of trust}: Blockchain-based PKIs based on WoT \cite{AliNSF16, AlBassamM17, QINHWLLS20, FromknechtVY14} typically replace CA with miners in public blockchains such as Bitcoin or Ethereum, and certificates are generated by mining after PoW. Certificate owners typically pay miners for their work and there is no mechanism for preventing or revoking mis-issued certificates. Another approach which is proposed in this paper is to distribute trust between entities, and certification and revocation take place after a consensus between entities. 
\end{itemize}}
\end{enumerate}
In this paper, we propose a dynamic blockchain-based PKI based on WoT and without any central role for the CA such that new keys can be securely enrolled or revoked based on a consensus mechanism between trusted nodes that are already part of the system. 
The possibility to dynamically add and revoke nodes is a requirement in fully-distributed and loosely-connected systems such as smart city and IoT applications. Furthermore, we seek a PKI solution for such systems such that despite the lack of a CA, the participating nodes can easily determine whether or not a particular public key can be regarded as trusted. 
This resolves a major problem in loosely-connected distributed systems where it is hard to assign a single or a few trusted nodes to enroll and revoke keys in the system. A PKI built on the assumption of the presence of such roles such as the one proposed in \cite{lewisonC16} will take away the main benefits of using a blockchain-based PKI to a large extent since enrollment and revocation decisions must always go through a traditional CA. Hence, we instead suggest a novel method where the key enrollment and revocation decisions are taken by the nodes that are part of the PKI and not a CA. This allows a sub-part of the system to act completely autonomous from the rest of the system for the key enrollment and revocation but comes at the price of the risk that a sufficient number of malicious nodes might be able to exclude honest nodes from the PKI. 
We resolve this problem by using the practical Byzantine fault tolerance (PBFT) \cite{CastroL99, CastroL02, Fan08, XuLLLG18} as the consensus mechanism within the enrollment and revocation procedures. The scheme will then tolerate up to $\lfloor(\thr-1)/3\rfloor$ malicious nodes where \thr denotes the number of nodes in the consensus group. 

Another problem in a blockchain-based PKI is the efficient verification of set membership. In a traditional PKI, such verification could be simply done by verifying the CA's signature in the certificate and verifying that the certificate is not revoked, for instance, using OCSP \cite{OCSP} or CRL \cite{rfc5280}. However, in an ordinary blockchain-based PKI where enrolled and revoked keys are added to the blockchain in chronological order, whenever a node wants to verify that a given public key is valid, it has to read many blocks on the chain and then verifies the corresponding signatures. This would be demanding, especially for resource-constrained nodes. To tackle this problem, we use a dynamic cryptographic accumulator, a space-efficient data structure that supports a fast verification of set membership. This can reduce the required time and space for verification from linear to logarithmic. Cryptographic accumulators were introduced by Benaloh and de Mare in 1993 as a decentralized alternative to digital signatures  \cite{BenalohM93} and have been used in different applications. When an element is added to a set through an accumulator, a witness is generated that can be used later to prove the membership of the element. We use a dynamic accumulator that supports the deletion of elements. Each new block in the proposed solution will contain an updated accumulator and witnesses. For verifying a key, a node will then only need to access the last block in the chain. The previous blocks are stored and available for auditing by anyone and further follow-up of malicious activities. Our contributions can be summarized as: 
\begin{itemize}
	\item We introduce a general model and a solution for dynamic blockchain-based PKI that we call it \scheme. It is decentralized, removes the need for having CAs and CRLs, and supports important functionalities for enrollment, revocation, update, and verification of public keys.
	\item We incorporate WoT into blockchain-based PKI such that enrollment and revocation of nodes are based on consensus between \thr number of nodes. We deploy PBFT as the consensus mechanism. 
	\item All the public keys are validated during the enrollment procedure. Any entity that is part of the system can act as an auditor and initiate the revocation procedure if it detects any malicious activity or invalid key. 
    \item For efficient verification of the public keys, we incorporate a \emph{dynamic} Merkle tree-based accumulator into the proposed solution. 
\end{itemize}

\subsection{Related work}
\label{sec:related}
Since the introduction of the Bitcoin \cite{nakamoto08}, blockchain technology received lots of attention and got many applications such as smart cities \cite{ShenP18} and Internet-of-Things (IoT) \cite{SinglaB18}. 
The blockchain enables secure and fully-distributed log management which is suitable for peer-to-peer networks. Transaction logs are included in a chain of blocks where each block contains a secure one-way hash of the previous block. A new block is only allowed to be added to the chain if it has gone through a \emph{consensus} decision between the peers in the network. 
The consensus mechanism is a critical part in blockchain and ensures its security and efficiency. The most utilized consensus mechanism in permissionless blockchain and cryptocurrencies is \emph{proof of work} (PoW), which is very energy consuming \cite{BachMZ18b}. 
\emph{Byzantine fault tolerant} (BFT) protocols are well-known consensus mechanisms that tolerate malicious faults, but many of them are inefficient to be used in practice or rely on assumptions that can be easily invalidated by an attacker. 

Most proposals for blockchain-based PKI simply consider the blockchain as an append-only public bulletin and keep having CA and thus provide very limited functionalities: Wang et al. \cite{WangLCWJZ18} used blockchain as an append-only public log to monitor CA's certificate signing and revocation operations on TLS certificates. CA-signed certificates and their revocation status are submitted by the web server as a transaction and appended to the blockchain by the miners after verifying transactions. This solution is still dependent on a traditional CA role, and the blockchain mainly serves as a source for certificate status registration and look-up. 
Kubilay et al. \cite{KUBILAYKM19} introduced CertLedger which uses public blockchain as a public log to validate, store, and revoke TLS certificates.  
Lewison et al. \cite{lewisonC16} used the Ethereum platform for PKI management and maintaining keys but used a centralized CA model for adding and revoking keys. The blockchain is then used as a public log, and the CA acts as the root of the chain and issues the certificates. 
Matsumoto and Reischuk proposed IKP \cite{MatsumotoR16}, an Ethereum-based PKI enhancement that uses smart contracts to offer automatic responses to CA misbehavior and incentives for those who help in detecting misbehaviors but still, CA has a central role and the trust model is hierarchical. 
Yang et al. \cite{yangSW18} proposed BC-PKM, a public key management system based on blockchain for named data networking. 
It includes basic functionalities that we consider in this paper but does not enjoy our WoT-based mechanism for adding and revoking public keys.
Qin et al. \cite{QINHWLLS20} proposed \emph{Cecoin} where digital certificates are treated as currencies and stored on a decentralized database where states are updated by miners on Bitcoin-based blockchain. They replace CAs with miners in the Bitcoin so for obtaining a certificate, the certificate owner must pay some cecoins to miners for their contributions and wait until they mine a new block through a PoW mechanism. They deploy a modified Merkle Patricia tree for retrieval and verification of certificates. 
AlBassam \cite{AlBassamM17} introduced SCPKI where each entity uses smart contracts to publish a set of attributes, signatures, and revocations for its identity on the Ethereum platform and allows entities to store, retrieve and verify identities through a PGP-like WoT. However, it does not provide functionalities and features considered in this paper and has other limitations including adaptability and privacy.
Ali et al. \cite{AliNSF16} introduced \emph{Blockstack} that leverages the Bitcoin blockchain to provide a name registration service that allows users to bind public keys to their names. Their solution was initially based on Namecoin, but after finding some security problems in Namecoin, they migrated to Bitcoin. \emph{Namecoin}, a cryptocurrency forked from the Bitcoin, was introduced for a decentralized DNS for \emph{.bit} addresses, where self-signed TLS certificates of a domain can be added to DNS addresses as auxiliary information. 
Fromknecht et al. \cite{FromknechtVY14} proposed \emph{Certcoin}, a blockchain-based PKI which ensures \emph{identity retention}, i.e. to prevent registering different public keys for one identity. Certcoin is based on Namecoin cryptocurrency and includes two versions: a version that supports efficient verification uses accumulators, and a version that supports efficient lookup is based on an authenticated distributed hash table. Certcoin is based on PoW and mining concept and does not enjoy our WoT and consensus mechanism or a dynamic accumulator that supports deletion. 
It cannot prevent identity squatting, does not support recoverability for identities that are falsely added to the blockchain, and does not fulfill a user's demand of using multiple public keys under the same identity, which is the case in many applications such as distributed IoT systems.
Anada et al. \cite{AnadaKWS14} proposed to include not only the subject ID into a public key value but also the guarantors' public key ID. They suggest using blockchain for maintaining the list of keys in a consistent way across the network. However, their approach does not cover the actual principles for maintaining this log but rather focuses on the inclusion of public key IDs into the public key values themselves.

Similar to some other blockchain-based PKI schemes \cite{AnadaKWS14, FromknechtVY14, lewisonC16}, our proposed scheme uses blockchain to maintain the list of all trusted and revoked public keys. 
However, different from previous work, in the proposed solution a new public key will be accepted and added to the blockchain if and only if the key is confirmed by at least a pre-defined number $\thr$ of trusted nodes that are already part of the system. This number will increase as the blockchain size increases. Similarly, keys can be revoked if at least $\thr$ trusted entities agree. Once a key is revoked, it is not allowed to be added to the chain again. Instead, units that want to join the PKI again will be forced to generate new key pairs and instantiate the enrollment procedure again. A comparison between the proposed scheme and related work is provided in Table \ref{table:relatedworks} where storage type denotes whether full or hash of public keys are stored on the chain. 

\begin{table*}[!t]
\centering
\resizebox{\textwidth}{!}{
\begin{tabular}{@{}cccccccccc@{}}
\toprule
Scheme & Centralization & Trust Model & Blockchain Type & Consensus & Certificate format &  Updatable key & On-chain storage \\ 
\midrule
CertLedger \cite{KUBILAYKM19} & Semi-centralized & Hierarchical & Ethereum-based & PBFT & X.509 & No & Hash only \\ 
Lewison et al. \cite{lewisonC16} & Semi-centralized & Hierarchical & Ethereum & N/A & Custom & No & Full \\ 
Wang et al. \cite{WangLCWJZ18} & Semi-centralized & Hierarchical & Custom & N/A & X.509 & Yes & Full \\  
Yakubov et al. \cite{YakubovSWSS18} & Semi-centralized & Hierarchical & Ethereum & N/A & X.509v3 & No & Full \\ 
CBPKI \cite{KhieuM19} & Semi-centralized & Hierarchical & Ethereum & N/A & X.509 & No & Hash only \\  
IKP \cite{MatsumotoR16} & Semi-centralized & Hierarchical & Ethereum-based & N/A & X.509 & Yes & Full  
\\ 
Blockstack \cite{AliNSF16} & Decentralized & WoT & Bitcoin & PoW & Custom & No & Hash only \\  
SCPKI \cite{AlBassamM17} & Decentralized & WoT & Ethereum & N/A & Custom & No & Hash only \\  
Cecoin \cite{QINHWLLS20} & Decentralized & WoT & Bitcoin & PoW & Custom & Yes & Full \\  
Certcoin \cite{FromknechtVY14} & Decentralized & WoT & Namecoin & PoW & Custom & Yes & Full \\ 
Proposed scheme (\scheme) & Decentralized & WoT & Custom & PBFT $^1$ & Custom & Yes & Hash only \\ \bottomrule 
\end{tabular}
}
\caption{Comparison between proposals for blockchain-based PKI \newline \footnotesize{\normalfont{$^1$ It can be replaced with more scalable and efficient variants of the PBFT.}}}
\label{table:relatedworks}
\end{table*}

We do not consider privacy-awareness \cite{AxonG17, OmololaP19} in this paper, but the \scheme can be updated to provide it if needed. Privacy-awareness in blockchain-based PKI was considered first in PB-PKI \cite{AxonG17} which is the same as the Certcoin in registering, verifying, and revoking public keys, but has a different key update mechanism. Each entity has offline and online pairs of public-private keys and registers its identity by posting its online public key on the blockchain. The main difference with the Certcoin resides in their key update procedure which removes the direct link between identity and public key and introduces an update function that given the new offline private key and old online public key, outputs a new online public key. The solution, however, is criticized in \cite{OmololaP19} by showing problems in the key update procedure, user authentication during the key update, and key revocation. 

\section{Preliminaries}
\label{sec:background}

\begin{definition}
	\label{def:signature}
	A \emph{signature scheme} $SIG = (\keygen, \sign, \veri)$ consists of the following polynomial-time algorithms: 
	\begin{itemize} 
		\item $\keygen(1^\lambda) \give (\sk, \pk)$: The key generation algorithm is a randomized algorithm that takes security parameter $\lambda$ as input and outputs a pair of private and public keys. 
		\item $\sign(\sk, \msg) \give \sigma$: The signing algorithm signs the message $\msg \in \{0,1\}^*$ with the private key $\sk$ and returns digital signature $\sigma$. 
		\item $\veri(\pk, \sigma, \msg) \give 1/\bot$: The signature verification is a deterministic algorithm that determines whether or not $\sigma$ is a valid signature for $\msg$ under the corresponding private key of public key $\pk$. 
	\end{itemize}
\end{definition}

\begin{definition}
	\label{def:hash}
	A hash function $\hash: \{0,1\}^* \xrightarrow{} \{0,1\}^n$ is cryptographic secure if it is preimage resistant, second preimage resistant, and collision resistant.
\end{definition}


\begin{definition}
	\label{def:eufcma}
	A signature scheme $SIG = (\keygen, \sign, \veri)$ is \emph{existentially unforgeable under an adaptive chosen message attack} or \eufcma-secure \cite{GoldwasserMR88} 
	if for any probabilistic polynomial-time (PPT) adversary \advA, the adversarial advantage in winning the security experiment defined in Figure \ref{fig:eufcma} is negligible, i.e. we have: 
	\begin{equation}
	    \Adv_{SIG}^{\eufcma} = \Pr[\Exp_{SIG}^{\eufcma}(\advA) = 1] \leq negl(\lambda)
	\end{equation}
	where $negl$ denotes a negligible function, and  $\lambda$ is the security parameter. 
\end{definition}

\begin{figure}[!t]
\begin{center}
    	\begin{multicols}{2}
		\begin{minipage}{0.5\linewidth}
			$\underline{\Exp_{SIG}^{\eufcma}(\advA)}$ \\
			$(\sk, \pk) \getsr \keygen(1^\lambda)$ \\
			$ML \gets \emptyset$ \\
			$(\sigma', m') \getsr \advA^{O.\sign}(\pk) $ \\
			$t \gets \veri(\pk, \sigma', m') \wedge (m' \notin ML)$ \\
			$\ret t$ 
		\end{minipage}
		\vfill
		\columnbreak
		\begin{minipage}{0.4\linewidth}  
			$\underline{O.\sign(m)}$ \\
			$\sigma \gets \sign(\sk, m)$\\
			$ML \getscup m $ \\ 
			$\ret (\sigma, m)$ 
		\end{minipage}
	\end{multicols}
	\caption{Security experiment for \eufcma-secure signature schemes. Adversary \advA can adaptively query $O.\sign$ oracle to sign its chosen messages.}
	\label{fig:eufcma}
\end{center}
\end{figure}

\begin{definition}
	A cryptographic \emph{accumulator} $ACC$ basically consists of four polynomial-time algorithms 
	\accgen, \accadd, \accwitadd and \accver \cite{ReyzinY16, BaldimtsiCDLRSY17}: 
	\begin{itemize} 
		\item $\accgen(1^\lambda) \give \acc_0$: given the security parameter $\lambda$, initiates the accumulator by an empty set $\acc_0$, as well as some additional parameters if needed. 
		
		\item $\accadd(\acc_i, x) \give (\acc_{i+1}, w^x_{i+1}, \upmsg_{i+1})$: takes in the current state of the accumulator $\acc_i$ and the value to be added $x$, and outputs an updated accumulator value $\acc_{i+1}$ and the membership witness $w^x_{i+1}$ for $x$. Additionally, an update message $\upmsg_{i+1}$ is generated  which can be used by any other witness holders to update their witnesses.
		
		\item $\accwitadd(w^x_i, y, \upmsg_{i+1}) \give w^x_{i+1}$: updates the witness for element $x$ after another element $y$ is added to the accumulator. The update message $\upmsg_{i+1}$ may contain any subset of $\{x, \acc_i, \acc_{i+1}, w^y_{i+1} \}$ and other parameters.
		
		\item $\accver(\acc_i, x, w^x_i) \give 1/\bot$: verifies the membership of $x$ in the accumulator using witness $w^x_i$ and accumulator state $\acc_i$. 
	\end{itemize} 
	
	Any accumulator should provide \emph{correctness} and \emph{soundness} properties. \emph{Dynamic} accumulators \cite{CamenischL02} that support deletion of elements from the accumulator have the following two additional algorithms: 
	\begin{itemize}
		\item $\accdel(\acc_i, x) \give (\acc_{i+1}, \upmsg_{i+1})$: deletes element $x$ from the accumulator. 
		
		\item $\accmemwitupondel(x, w^x_i, \upmsg_{i+1}) \give w^x_{i+1}$: After deletion of $y$ from the accumulator, it updates the membership witness for element $x$. 
	\end{itemize}
\end{definition}

\begin{definition}
	\label{def:accsound}
	A dynamic accumulator is \emph{sound} (or simply secure) if it is difficult to fabricate a witness $w^x$ for a value $x$ that has not been added to the accumulator \cite{ReyzinY16, BaldimtsiCDLRSY17}. More formally, for any security parameter $\lambda$ and any stateful PPT adversary \advA with black-box access to \accadd and \accdel oracles which take elements $x_0$ on accumulator \acc, we should have:
	\begin{align}
	\Pr \left[ \begin{array}{l}
	\acc_0 \gets \accgen(1^\lambda); L \gets \emptyset;\\
	(x, w^x) \gets \advA^{\accadd,\accdel(\acc_0, x_0)}; \\
	x \notin L: 
	\accver (\acc_1, x, w^x) = 1 \\
	\end{array}\right] \leq negl(\lambda)
	\end{align}
	where $negl$ is a negligible function in the security parameter $\lambda$, and $\acc_1$ denotes the accumulator state at the end of the security experiment. 
	$L$ (which is initiated with an empty set) keeps the list of elements that are used in adversarial calls, and will be updated after each adversarial call to the \accadd and \accdel oracles. 
\end{definition}

\begin{definition}
	We define a \emph{ledger} $\mathbf{x} \in \ledger$ as a vector of sequences of transactions $\tx \in \mathcal{T}$. 
	Transactions are defined through a digital signature scheme as defined in Definition \ref{def:signature}. 
	A transaction \tx is of the form $\{\id_1, \id_2, \dots, \id_\thr\} \rightarrow (\sigma, \{ \} )$ where $\sigma$ is a vector $\langle (\pk_1, \sigma_1), \dots, (\pk_\thr, \sigma_\thr)\rangle$ of public keys and corresponding digital signatures. 
	A transaction \tx is valid with respect to a ledger 
	if all digital signatures are verified. 
\end{definition}

\section{PBFT}
\label{sec:PBFT}
\begin{figure}[!t]
    \centering
    \includegraphics[width=0.63\textwidth]{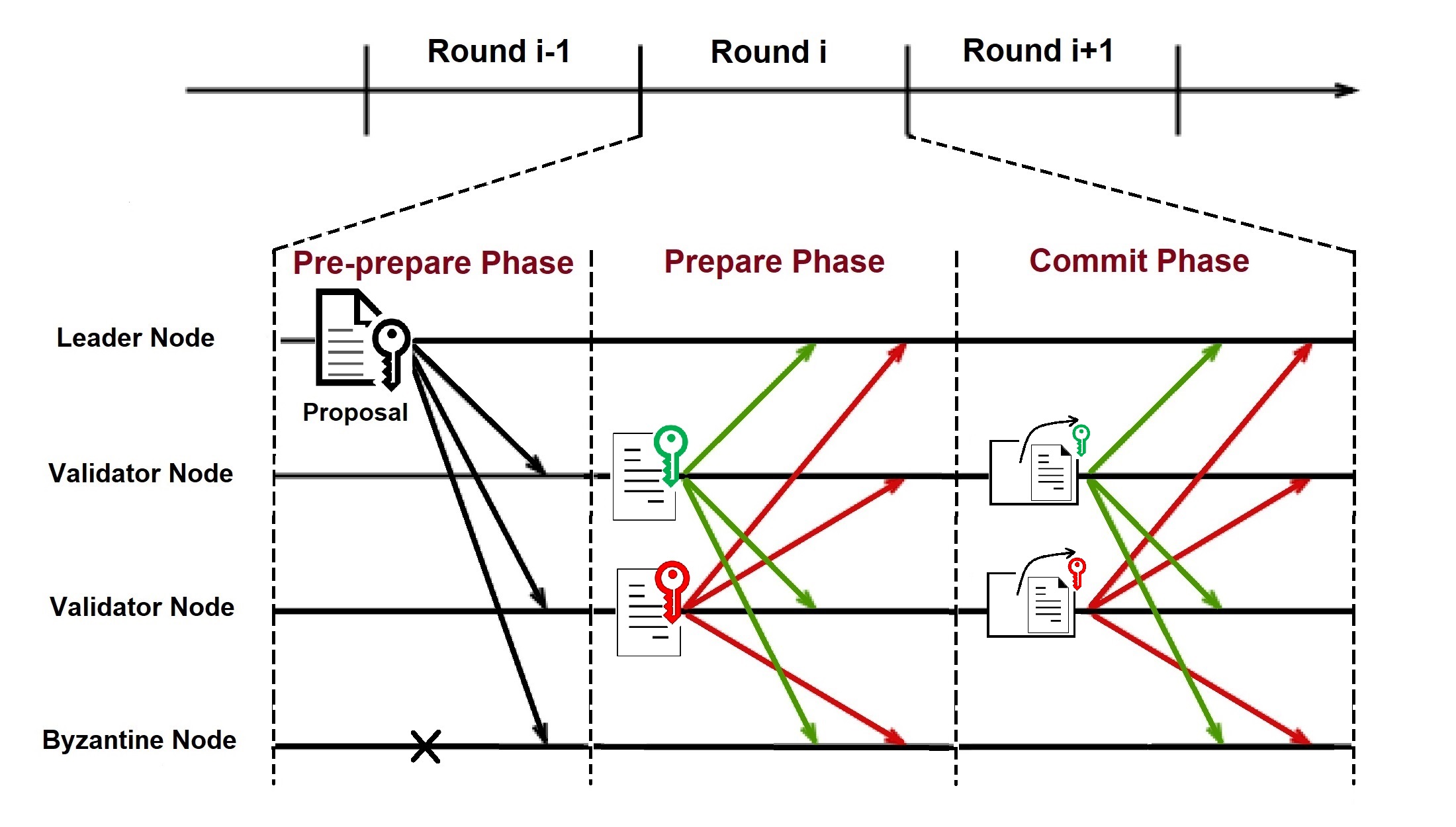}
    \caption{Overview of different phases in PBFT \cite{CastroL99, CastroL02, Fan08}.}
    \label{fig:PBFT}
\end{figure}

PBFT \cite{CastroL99, CastroL02} was the first practical and efficient solution to deal with Byzantine faults in a weakly synchronous environment. It executes in some rounds between some nodes in a consensus group in which one node acts as the \emph{leader}, and other nodes act as \emph{validators}. As depicted in Figure \ref{fig:PBFT}, all messages are signed before broadcasting to other nodes in the consensus group. Each round consists of the following phases:
\begin{enumerate}
\item \emph{Pre-prepare phase}: The leader initiates the consensus procedure by sending a signed \preprepare message that is a block proposal containing a certain number of transactions.
\item \emph{Prepare phase}: Upon receiving the \preprepare message, each node in the consensus group checks the correctness and validity of the block and multicasts a signed \prepare message (yes/no) to all other nodes.
\item \emph{Commit phase}: Each node, after analysis of received \prepare messages, multicasts a signed \commit message (yes/no) to the consensus group. The block proposal is committed to the blockchain only if a sufficient number of nodes in the consensus group agree on it.
\end{enumerate}

At the end of the above phases, all honest nodes in the consensus group reach a consensus about accepting or rejecting the block proposal and will have the same view about the blockchain state. 

Since a consensus group of \thr players can tolerate $f=\lfloor(\thr-1)/3\rfloor$ Byzantine nodes in the PBFT, each honest node needs to collect and verify at least $2f + 1$ signatures in both prepare and commit phases, respectively. This might limit the application of PBFT to small consensus groups due to communication and computational overhead, but there are proposals for improving the scalability of PBFT and reducing its computational costs \cite{Fan08, XuLLLG18, MirBFT}. 
As it will be discussed in Section \ref{sec:discussion}, recent proposals for BFT such as HotStuff \cite{YinMRGA19} (and its variant LibraBFT \cite{Baudet19state}) are more efficient but new proposals would need more scrutiny \cite{Momose19} so we would prefer to deploy the well-studied PBFT in this paper. 

\section{Security Model}
\label{sec:model}

Based on basic functionalities that we consider for a PKI which includes registering, revoking, and updating public keys, an adversary might perform the following attacks:  

\begin{itemize}
    \item Registering a valid public key for an illegitimate entity
    \item Registering an invalid public key for an illegitimate entity
    \item Registering an invalid public key for a legitimate entity
    \item Updating public key of a legitimate entity with a valid/invalid public key
    \item Revoking public key of a legitimate entity
\end{itemize}

Validation of public keys is very important and invalid public keys can help an adversary to accomplish some attacks: 
In protocols based on the discrete logarithm problem, a small-subgroup attack might be feasible if public keys are not validated to be of prime order. A variant of this attack in elliptic curve-based schemes is the invalid-curve attack \cite{AntipaBMSV03, Toorani16}. We formalize such validations through the following definition. 

\begin{definition}
We denote by \emph{Public key and Identity Validation} (\pkveri) the procedure taken in the PKI for validation of public keys and identifiers: 
\begin{equation*}
\pkveri(\id, \pk) \give 1/\bot  
\end{equation*}
We denote by $\regpkspace$ and $\regidspace$ the list of all public keys and valid identifiers that are already registered in a PKI, respectively. We denote by $\pkspace$ the space of all valid public keys generated by the $\keygen(1^\lambda)$ algorithm using the randomness $r \in_R \{0,1\}^\lambda$: \\
\begin{equation*}
\pkspace = \{ \pk | \exists r \in \{0,1\}^\lambda:    \keygen(1^\lambda;r)=(\sk,\pk)\} 
\end{equation*}
\end{definition}

For a PKI that does not check the validity of public keys and identifiers, we have $\pkveri(.,.)=1$.
A PKI may deploy zero-knowledge proofs through a so-called \emph{proof-of-possession} (PoP) procedure to assure that any user owns the corresponding private key of its claimed public key. Although standards mandate the inclusion of PoP during registration, many existing PKIs do not require proofs of knowledge \cite{RistenpartY07}. 

\begin{remark}
For a PKI that only considers \emph{public key validation}, we define \pkveri as: 
\begin{equation*}
    \pkveri(\id, \pk) = 
    \begin{cases}
      1    & \text{if}\ \pk \in \pkspace \\
      \bot & \text{otherwise}
    \end{cases}
\end{equation*}
\end{remark}

\begin{remark}
For providing \emph{public key uniqueness} in a PKI, we define \pkveri as: 
\begin{equation*}
    \pkveri(\id, \pk) = 
    \begin{cases}
      1    & \text{if}\ (\pk \not\in \regpkspace) \wedge (\pk \in \pkspace) \\
      \bot & \text{otherwise}
    \end{cases}
\end{equation*}
\end{remark}

\begin{remark}
For preserving \emph{identity retention} (preventing different public keys registered for one identifier) in a PKI, we define \pkveri as: 
\begin{equation*}
    \pkveri(\id, \pk) = 
    \begin{cases}
      1    & \text{if}\ (\id \not\in \regidspace) \wedge (\pk \in \pkspace) \\
      \bot & \text{otherwise}
    \end{cases}
\end{equation*}
\end{remark}

We will use \pkveri to provide a generic model for PKI and to formalize the security definitions for an adversary that aims to register valid and invalid public keys for different identifiers of her will. 

\section{DBPKI}
\label{sec:scheme}

The proposed scheme \scheme includes the following entities: 
\begin{itemize}
    \item \emph{Root units} ($\ur_i$): Each root entity is part of the \scheme and assumed to be honest in the beginning. It is identified by an identifier $\id_{\ur_i}$, and has a pair of private and public keys $(\sk_{\ur_i}, \pk_{\ur_i})$ and a fixed trust weight $v_r$. 
    \scheme is initialized by $n$ root units $\{\ur_1, \dots, \ur_n \}$. 

    \item \emph{Intermediate units} ($\ui_i$): Each intermediate entity is part of the \scheme, identified by an identifier $\id_{\ui_i}$, and has a pair of private and public keys $(\sk_{\ui_i}, \pk_{\ui_i})$ and a fixed trust weight 1. 
    Intermediate units could be organizations in applications like smart cities, and their number can be increased dynamically per se. 
        
    \item \emph{Ordinary units} ($\uo_i$): Each ordinary entity is \emph{not} part of the \scheme. It is identified by an identifier $\id_{\uo_i}$, and has a pair of private and public keys $(\sk_{\uo_i}, \pk_{\uo_i})$ and a fixed trust weight \emph{zero}. This means that they cannot participate in enrollment and revocation procedures, but will be ordinary users of the system. They might be resource-constrained units that will join and use the system but will not be part of the \scheme. 
\end{itemize}

When referring to an entity in general, we denote it by $u_i$ which belongs to one of above groups and has a $\role\in\{\ur, \ui, \uo\}$. 
Only root and intermediate units can participate in enrollment, revocation, and update procedures while any node can accomplish the verification procedure as described in Section \ref{sec:functionalities}. 
Any entity which is part the \scheme, i.e. with  $\role\in\{\ur, \ui\}$, can initiate enrollment, revocation, or update procedures by establishing the PBFT consensus mechanism 
as a \emph{leader} with  $\thr-1$ number of other units that are part of the \scheme. The block proposal will be added to the blockchain only if the consensus is achieved. 

An \emph{account} is identified by $(\id_{i}, \hash(\pk_{u_i}), \role)$ where $\id_i$ is an identifier that uniquely identifies the account. 
Any entity $u_i$ may have one or more accounts. By incorporating the public key validation function \pkveri into the \scheme, only valid public keys will be accepted. This prevents further attacks and can support different policies such as public key uniqueness or identity retention.

\subsection{Blockchain structure}
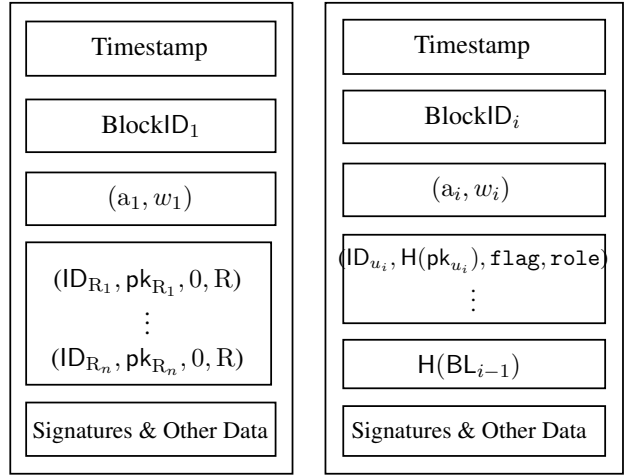
\begin{figure}
    \centering
\tikzset{every picture/.style={line width=0.75pt}} 
\resizebox{0.5\linewidth}{!}{
\begin{tikzpicture}[x=0.75pt,y=0.75pt,yscale=-1,xscale=1]
\draw   (238,20.45) -- (405,20.45) -- (405,285) -- (238,285) -- cycle ;
\draw   (61.09,20.45) -- (219.51,20.45) -- (219.51,285) -- (61.09,285) -- cycle ;
\draw   (248,29.95) -- (395,29.95) -- (395,60.45) -- (248,60.45) -- cycle ;
\draw   (248,69.75) -- (395,69.75) -- (395,99.35) -- (248,99.35) -- cycle ;
\draw   (248,110.75) -- (395,110.75) -- (395,140.35) -- (248,140.35) -- cycle ;
\draw   (248, 150.75) -- (395,150.75) -- (395,200.45) -- (248,200.45) -- cycle ;
\draw   (248,209.75) -- (395,209.75) -- (395,237) -- (248,237) -- cycle ;
\draw   (248,247) -- (395,247) -- (395,275) -- (248,275) -- cycle ;
\draw   (70,30.95) -- (210,30.95) -- (210,61.45) -- (70,61.45) -- cycle ;
\draw   (70,74.75) -- (210,74.75) -- (210,104.35) -- (70,104.35) -- cycle ;
\draw   (70,115.75) -- (210,115.75) -- (210,145.35) -- (70,145.35) -- cycle ;
\draw   (70,155) -- (207.51,155) -- (207.51,235) -- (70,235) -- cycle ;
\draw   (70,245) -- (210,245) -- (210,275) -- (70,275) -- cycle ;
\draw (320.3,45.2) node  
[align=center] {Timestamp};
\draw (320.63,84.55) node  
[align=center] {Block$\id_i$};
\draw (319.63,125.55) node  
[align=center] {$(\acc_i, w_i)$};
\draw (321,175.6) node  [font=\small] 
[align=center] {($\id_{u_i}, \hash(\pk_{u_i}) ,\flag, \role$)\\$\vdots$};
\draw (318.63,224.55) node  
[align=center] {$\hash(\blockchain_{i-1})$};
\draw (318.63,262) node  [font=\small]
[align=center] {Signatures \& Other Data};
\draw (139.3,46.2) node   
[align=center] {Timestamp};
\draw (140.13,89.55) node  
[align=center] {Block$\id_1$};
\draw (139.63,130.55) node  
[align=center] {$(\acc_1, w_1)$};
\draw (138.13,199.6) node   
[align=center] {($\id_{\ur_1},\pk_{\ur_1}, 0, \ur$)\\ $\vdots$ \\ ($\id_{\ur_n},\pk_{\ur_n}, 0, \ur$)};
\draw (139.63,262) node  [font=\small]
[align=center] {Signatures \& Other Data};
\end{tikzpicture}}
    \caption{Contents of the first block $\blockchain_1$ (left) and $i^{th}$ block $\blockchain_i$ (right).}
    \label{fig:blockchain}
\end{figure}

Items that will be included in the blockchain are listed in Figure \ref{fig:blockchain} where $\role \in \{\ur,\ui,\uo\}$ denotes the role of entity $u_i$ with identifier $\id_{u_i}$, and $\flag \in \{0,1,2\}$ denotes that public key $\pk_{u_i}$ is newly added, updated, or revoked, respectively. 
The very first block in the chain contains public keys and identifiers of all $n$ root units: $\{(\id_{\ur_1}, \pk_{\ur_1}, 0, \ur), \dots, (\id_{\ur_n}, \pk_{\ur_n}, 0 , \ur)\}$ together with a timestamp, block identifier, accumulator, and some additional data. 
The next blocks in the chain, as depicted in Figure \ref{fig:blockchain}, include a hash of public key instead of the public key itself. This will decrease the blockchain size. 
An arbitrary block $\blockchain_i$ may contain an arbitrary \nrTI number of key transaction items. 
Each valid block $\blockchain_i$ contains a secure one-way hash of the previous block in the chain $\hash(\blockchain_{i-1})$ where $i$ denotes the index of the current block. 
A single key transaction item includes adding, updating or revocation of public key of a unit $u_i$. A key transaction item is valid and will be added to the blockchain only after that a consensus is achieved between \thr number of units that are already part of the \scheme. 
Any efficient and sound dynamic Merkle-tree based accumulator such as those proposed in \cite{ReyzinY16, BaldimtsiCDLRSY17} can be used in the scheme. 

\subsection{Functionalities}
\label{sec:functionalities}
The proposed scheme provides the following functionalities as defined in Figure \ref{fig:functions}:
\begin{itemize}
	\item $\register (\id_{u_i}, \pk_{u_i}, \flag, \role) \give 1/\bot$: Enrolls an entity with $\role\in\{\ur, \ui, \uo\}$, identifier $\id_{u_i}$, and public key $\pk_{u_i}$, and outputs success (1) or failure ($\bot$). Here, $\flag$ can be either 0 or 1 where 0 means $\pk_{u_i}$ has been registered for the first time for $\id_{u_i}$, and 1 means the public key has been updated due to an update procedure. 
	
	\item $\revoke(\id_{u_i}, \pk_{u_i}) \give 1/\bot$: Revokes the public key $\pk_{u_i}$ corresponding to identity $\id_{u_i}$, and outputs success (1) or failure ($\bot$).

	\item $\update(\id_{u_i}, \pk^*_{u_i}) \give 1/\bot$: Updates the corresponding public key of an entity with identifier $\id_{u_i}$ to $\pk^*_{u_i}$, and outputs success (1) or failure ($\bot$). 

	\item $\verify(\id_{u_i}, \pk_{u_i}) \give 1/\bot$: Verifies whether or not $\pk_{u_i}$ is the corresponding public key of $\id_{u_i}$, and outputs success (1) or failure ($\bot$).
\end{itemize}

\begin{figure}[!t]
	\setlength{\columnseprule}{0.35pt}
	\begin{multicols}{2}
		\begin{minipage}{\linewidth}
			$\underline{\setup(\lambda)}$ \\
			$\templist \gets\emptyset$ \\
			$\acc_0 \gets \accgen(1^\lambda)$ \\
			$\foor i \in\{1,\dots,n\} ~\doo$ \\
			$\dent(\sk_{u_i}, \pk_{u_i}) \gets \keygen(1^\lambda)$ \\
			$\dent\templist\getscup(\id_{R_i}, \pk_{R_i}, 0, \ur)$ \\
			$\dent x \gets (\id_{R_i}, \hash(\pk_{R_i}))$ \\
			$\dent\accadd(\acc_0, x)$ \\
            $\addBL(\{Opt, \acc_1, w_1, \templist \})$ \\

			$\underline{\addBL(Proposal)}$ \\
			\iif consensus achieved\\ 
			$\dent\blockchain_{i+1} \getscup \{Proposal\}$ \\

			$\underline{\register(\id, \pk, \flag, \role)}$ \\
			$\iif \pkveri(\id, \pk)=1 \wedge \\ (|SI|_{valid} \geq 2f+1)  
			~\then$ \\
			$\dent\templist\gets(\id, \hash(\pk), 0, \role)$ \\
			$\dent x \gets (\id, \hash(\pk))$ \\
			$\dent (\acc_{i+1},w_{i+1}) \gets \accadd(\acc_i, x)$ \\
			$\dent\addBL(\{Opt, \acc_{i+1}, w_{i+1}, \templist\})$ \\
			$\dent\ret 1$	\\
			$\eelse$ \\
		    $\dent\ret\bot$ 
			\end{minipage}
			\vfill
			\columnbreak
			\hspace{11pt}
		\begin{minipage}{\linewidth}
            $\underline{\verify(\id, \pk)}$ \\
			$x \gets (\id, \hash(\pk))$ \\ 
			$\iif \accver(\acc_i, x, w_i) = 1$\\
			$~\then ~\ret 1$ \\
			$\eelse$ \\
		    $\dent\ret\bot$ \\

			$\underline{\revoke(\id, \pk)}$ \\
		    $\iif \verify(\id,\pk) \wedge \\ 
		    (|SI|_{valid} \geq 2f+1) ~\then$ \\
			$\dent\accdel(\acc, (\id, \pk))$ \\
			$\dent\addBL(\id, \pk, 2)$ \\
		    $\eelse$ \\
	        $\dent\ret\bot$ \\

			$\underline{\update(\id, \pk^{old}, \pk^{new})}$ \\
			$\iif \pkveri(\id, \pk^{new})=1 \then$ \\
			$\dent\revoke(\id, \pk^{old})$ \\
			$\dent\register(\id, \pk^{new}, 1, \role)$ \\
			$\dent\ret~1$ \\
			$\eelse$ \\
		    $\dent\ret\bot$
		\end{minipage}
	\end{multicols}
	\caption[Functionalities]{Definitions for basic functionalities in \scheme.} 
	\label{fig:functions}
\end{figure}

The enrollment, revocation, and update procedures require consensus between at least $\thr-1$ other units that are already part of the \scheme where $\thr$ is a system dependent parameter. 
To facilitate the election procedure within the consensus, \thr number of roots or intermediate nodes are assigned as members of the \emph{consensus group}. The consensus group will be updated if some units are revoked at a later time or if the threshold value \thr is changed. PBFT is used as the consensus algorithm. 

\subsubsection{Initialization}
\label{sec:init}
\begin{enumerate}
    \item Each of $n$ root units $\{\ur_1, \dots, \ur_n\}$ generates a key pair $(\sk_{\ur_i}, \pk_{\ur_i}) \gets \keygen(1^\lambda)$.
        
    \item Unit $\ur_1$ creates the accumulator $\acc_0 \gets \accgen(1^\lambda)$. It then adds identifiers and hash of public keys for all root units $(\id_{\ur_i}, \hash(\pk_{\ur_i}))$ to the accumulator where $i \in \{1, \dots, n\}$. The accumulator and witnesses are denoted by $\acc_1$ and $w_1$ after those updates. It then generates a block proposal $\blockchain_1$ which includes timestamp, Block$\id_1$, $\acc_1$, $w_1$, and ${(\id_{\ur_i}, \pk_{\ur_i}, 0, \ur)}$ for $i\in\{1,\dots,n\}$. 
    Then, $\ur_1$ initiates the PBFT protocol as the leader and proceeds the consensus mechanism with the other root units. 
    
    \item All root units participate in the BPFT protocol and during the consensus procedure, verify that the accumulator is created correctly. At the end of the consensus procedure, they reach to agreement on $\blockchain_1$ (and $\acc_1$). 
    Otherwise, the initialization procedure starts from the beginning. 
\end{enumerate}

\subsubsection{Enrollment procedure} 
\label{sec:enrollment}
Any arbitrary unit $u_i$ that wants to join the \scheme needs to generate a private-public key pair ($\sk_{u_i}$, $\pk_{u_i}$). 
Then, $u_i$ can initiate the enrollment procedure by sending a enrollment request to another unit $u_j$ which is in the consensus group. 
The enrollment request includes a data structure that we refer to as \emph{Enrollment Proof (\EP)}, and a \emph{public key item} (\PI). \EP indicates specifications of the enrollment (and further information when the unit has already a certificate from another trusted PKI as will be discussed in Section \ref{sec:discussion}).  
\PI includes $\{\id_{u_i}, \hash(\pk_{u_i}), \flag, \role, \Opt\}$ in which \Opt denotes additional data such as requested validity period of the key if applicable, subject information, etc. 
Then, any unit $u_j$ in the consensus group verifies the claimed public key and checks if $u_i$ is allowed to join the \scheme. Then, $u_j$ signs the proposal with its own private key $\sk_{u_j}$ and proceeds with the PBFT consensus mechanism with the other units in the consensus group. The block proposal will be added to the ledger only after a successful consensus. 
Steps for the enrollment procedure can be followed as:

\begin{enumerate}
\item $u_i$ generates a key pair $(\sk_{u_i}, \pk_{u_i}) \gets \keygen(1^\lambda)$. It then generates an enrollment proof \EP, a \emph{public key item} $\PI \gets \{\id_{u_i}, \hash(\pk_{u_i}), \flag, \role, \Opt\}$, 
a \emph{signature item} $\SI \gets \sign(\sk_{u_i}, \PI)$, and an \emph{enrollment request} which contains $\{\EP, \PI, \SI, \pk_{u_i}\}$ and will be sent to the consensus group. 

\item Any unit $u_j$ in the consensus group that has received an enrollment request $\{\EP, \PI, \SI, \pk_{u_i} \}$ and trusts $u_i$, checks that \flag and \role have correct values and verifies the signature included in the enrollment request. If $\veri(\sk_{u_i}, \SI, \PI) = 1$ then $u_j$ verifies that the public key $\pk_{u_i}$ is valid: If $\pkveri(ID_{u_i}, \pk_{u_i}) = 1$ then $u_j$ signs the received $\PI$ using its private key $\sk_{u_j}$, generates another \emph{signature item} $\SI \gets \sign(\sk_{u_j},\PI)$, broadcasts $\{\PI, \SI\}$ to the other units in the consensus group, and proceeds according to steps in the PBFT consensus mechanism. 

\item During the consensus procedure, any unit in the consensus group verifies that any signature item \SI from other units is valid and its signer is part of the \scheme. If any signature is not verified, the corresponding pair of $\{\PI, \SI\}$ will be discarded. Upon successful completion of the consensus mechanism in previous step, which means there are at least $2\lfloor(\thr-1)/3\rfloor+1$ number of valid $\{\PI,\SI\}$ for a particular \PI, all the honest units in the consensus group accept the proposal for enrollment of $u_i$ and have the same view about the blockchain state. 
All the units in the consensus group also add $x =(\id_{u_i}, \hash(\pk_{u_i}))$ to the accumulator: $(\acc_{i+1}, w^x_{i+1}, \upmsg_{i+1}) \gets \accadd(\acc_i, x)$. 
The new block $\blockchain_m$ that will be added to the ledger contains a key transaction item $\TI$ which includes \PI and all corresponding $\SI$s during the consensus procedure together with the updated accumulator and witness. $\blockchain_m$ will be broadcast to the whole network.

\item All block recipients verify that the accumulator has been updated correctly. If verified, they update their stored accumulator values with $\acc$, and update their stored witnesses as $w^x_{i+1} \gets \accwitadd(w^x_i, y, \upmsg_{i+1})$ where $y$ denotes the newly added pairs of identifiers and public keys. Otherwise, they discard the block or initiate the revocation procedure. 
\end{enumerate}

\subsubsection{Revocation procedure}
\label{sec:revocation}
Any arbitrary entity $u_i$ which is part of the \scheme can 
initiate the revocation procedure.  
The revocation procedure starts when $u_i$ finds a reason to revoke an existing key of entity $u_t$ in the ledger. 
This can happen due to bad or malicious behavior, due to that the key is obsolete, upon request from the owner of a key when its private key has been compromised, or other reasons. 
Entity $u_i$ sends a revocation request, and broadcasts it to the consensus group. The revocation request includes a data structure that we refer to it as \emph{Revocation Proof (\RP)} which indicates a reason for the revocation request. 
Any unit in the consensus group, checks \RP before further processing of the revocation request. Steps for the revocation procedure are as follows:

\begin{enumerate}
    \item $u_i$ generates a revocation proof \RP, a \emph{public key item} (\PI) as $\PI \gets \{\id_{u_i}, \hash(\pk_{u_i}), 2, \role, \Opt\}$ where 2 denotes the \flag for revocation, and a signature item $\SI \gets \sign(\sk_{u_i}, \PI)$. It then sends the revocation request $\{\RP, \PI, \SI, \pk_{u_i}\}$ to the consensus group which includes at least $\thr-1$ other entities that are already part of the \scheme and might accept revoking the public key of $u_t$. 
    
    \item Any node $u_j$ in the consensus group that has received the revocation request checks the received \RP and verifies the signature item \SI. If both are verified, $u_j$ signs the received \PI with its private key $\sk_{u_j}$ and generates a signature item $\SI \gets \sign(\sk_{u_j}, \PI)$, and broadcasts $\{\PI, \SI\}$ to other units in the consensus group, and proceeds according to steps in the PBFT consensus mechanism. 
    
    \item During the consensus procedure, any unit in the consensus group verifies that any signature item \SI from other units is valid and its signer is part of the \scheme. If any signature is not verified, the corresponding pair of $\{\PI, \SI\}$ will be excluded from further consideration. Upon successful completion of the consensus mechanism in previous step which means there are at least $2\lfloor(\thr-1)/3\rfloor+1$ number of valid $\{\PI,\SI\}$ for a particular \PI, all the honest units in the consensus group accept the proposal for revoking $u_i$ and have the same view about the blockchain state. All the units in the consensus group also include \PI and all valid $\SI$s together with identifiers and hash of public keys of the signing entities into a new key transaction item $\TI$. They also verify that $x = (\id_{u_t}, \hash(\pk_{u_t}))$ exists in the accumulator by checking that $\accver(\acc_i, x, w^x_i) = 1$. If the verification fails, the procedure aborts. Otherwise, they delete $(\id_{u_t}, \hash(\pk_{u_t}))$ from the accumulator by executing $(\acc_{i+1}, \upmsg_{i+1}) \gets \accdel(\acc_i, x)$, and append the updated accumulator $\acc_{i+1}$ to \TI. The new block $\blockchain_m$ that will be added to the ledger contains a key transaction item $\TI$ which includes \PI and all corresponding $\SI$s during the consensus procedure, together with the updated accumulator and witnesses. $\blockchain_m$ will be broadcast to the whole network. 
    
    \item All block recipients verify that the accumulator has been updated correctly. If verified, they update their stored accumulator values with $\acc$, and update their stored witnesses as $w^x_{i+1} \gets \accmemwitupondel(x, w^x_i, \upmsg_{i+1})$. Otherwise, they discard the block. 
\end{enumerate} 

\subsubsection{Update procedure}
\label{sec:update}
The update procedure should naturally include revoking the old key $\pk^{old}_{u_i}$ by executing $\revoke(\id_{u_i}, \pk^{old}_{u_i})$ and registering the new key $\pk^{new}_{u_i}$ 
by executing $\register(\id_{u_i}, \pk^{new}_{u_i}, 1, \role)$ where $\flag=1$ means that there are previous records for the public key in the ledger and it has been updated.
Since \register and \revoke procedures require consensus between \thr nodes, for the sake of efficiency, both procedures will be merged, i.e. the block proposal in the consensus mechanism includes records for revoking $\pk^{old}_{u_i}$ and enrollment of $\pk^{new}_{u_i}$. The new block that will be added to the ledger contains $(\id_{u_i}, \hash(\pk^{old}_{u_i}), 2, \role)$ and $(\id_{u_i}, \hash(\pk^{new}_{u_i}), 1, \role)$. 

\subsubsection{Verification procedure}
\label{sec:verification}
The verification procedure is essentially a proof of membership algorithm and is used to verify whether or not a given public key belongs to a given identifier. It naturally considers the revocation status of a key and will respond with true if and only if the public key is valid and belongs to the given identifier. This can decrease the computational costs, especially for resource-constrained environments.  
For a given $x=(\id_{u_j}, \hash(\pk_{u_j}))$, any node can accomplish the verification procedure by verifying that $\accver(\acc_i, x, w^x) = 1$. 

\section{Further discussion}
\label{sec:discussion} 

The computational and communication costs of the \scheme depend mostly on the deployed dynamic accumulator and consensus mechanism. Table \ref{table:costs1} shows the computational costs of different procedures in the \scheme. Computational costs of different accumulator's algorithms in some constructions are shown in Table \ref{table:costs2}. 
Accumulators with low-frequency updates require fewer updates which decreases the communications overhead. 
The consensus procedure is one of the costly parts and its costs depend on the number of nodes in the consensus group \thr and the deployed algorithm. 
There is a trade-off between security and efficiency. Increasing \thr will naturally increase security (since to withstand against $f$ corrupt nodes, we need to have at least $\thr = 3f+1$ nodes in the consensus group) but decreases efficiency (as computational costs and communications overhead will increase). 
PBFT is not the most efficient BFT algorithm but is well-studied. PBFT is safe against $f$ Byzantine faults, safe against asynchrony, and responsive (i.e., a leader node can propose without delay). However, the number of messages for a consensus decision and the number of messages to rotate a leader are both quadratic. There were also other options that we could deploy: Mir-BFT \cite{MirBFT} is a recent proposal based on the PBFT that allows having multiple leaders instead of one leader and improves the throughput by requiring fewer signatures. 
HotStuff \cite{YinMRGA19} and its underdeveloped variant, LibraBFT \cite{Baudet19state}, are other proposals that claim to provide the same properties as the PBFT but with a linear number of messages for a consensus decision and rotating a leader. 
We could incorporate those new BFT protocols into the scheme to make it more efficient and scalable but we simply used a classical BFT in this paper since new proposals would require further scrutiny and some of them have security vulnerabilities \cite{Momose19}. 

\begin{table*}[!t]
	\begin{center} 
		\begin{tabular}{@{}ccccccc@{}}
			\toprule 
			Procedure & \sign /\veri & \accadd & $\accwitadd^*$ & $\accmemwitupondel^*$ & \accdel & \accver \\ 
			\midrule
			\register & $O(\thr^2)$ & $\thr$ & $k$ & - & - & - \\  
			\revoke   & $O(\thr^2)$ & - & - & $k$ & \thr & \thr \\ 
			\update & $O(\thr^2)$ & \thr & $k$ & $k$ & \thr & \thr \\ 
			\verify & - & - & - & - & - & 1 \\ \bottomrule 
		\end{tabular}
	\end{center} 
	\caption{Total computational costs of different procedures in \scheme where \thr denotes the number of nodes in the consensus group, and $k$ denotes the number of nodes that store and validate the accumulator and witnesses.}
	\label{table:costs1}
\end{table*}

\begin{table*}[!t]
	\begin{center} 
		\begin{tabular}{@{}cccccccc@{}}
			\toprule 
			Scheme & \sign & Merkle & BraavosB \cite{BaldimtsiCDLRSY17} & CL-RSA-B \cite{BaldimtsiCDLRSY17} & Braavos \cite{BaldimtsiCDLRSY17} \\ 
			\midrule
			\accadd & 1 & $\log d$ & 1 & 1 & 1 \\  
			\accwitadd & 0 & $\log d$ & 0 & 0 & 0 \\  
			\accmemwitupondel & 0 & $\log d$ & 1 & 1 & 1 \\  
			\accdel & 0 & $\log d$ & 1 & 1 & 1 \\  
			\accver & 1 & $\log d$ & 1 & 1 & 1 \\  \bottomrule 
		\end{tabular}
	\end{center} 
	\caption{Costs of incorporating different accumulators into procedures where $d$ denotes the number of elements added to the accumulator.}
	\label{table:costs2}
\end{table*}

Other alternatives to the proposed scheme could also be considered: A variant that could reduce the number of required signatures during the enrollment and revocation procedures is to incorporate the trust weights in reaching the threshold. Since the trust weight of root and intermediate units is defined as $w_r$ and 1, respectively, each signature from a root entity could worth like $w_r$ signatures from intermediate entities. This reduces the waiting time which could be useful in emergency cases but it requires some changes in the consensus mechanism.

Another concern regarding the proposed solution is the interoperability with the existing CA-based PKI. This could be done by establishing a trust policy where members of the consensus group would consider certificates issued by some CAs as trusted. Then, any unit having a certificate from a trusted CA can include such specification in the enrollment proof \EP that is generated through the enrollment procedure. 
An example case would be a vehicle holding a certificate from a CA in another country that wants to visit a smart city deploying the \scheme. However, this should be a temporary enrollment and the unit shall join as an ordinary unit especially if the other CA-based PKI does not provide certificate transparency. 

In the rest of this section, we briefly argue about the security of the \scheme. Our security definitions follow a provable security game-based approach. For the security model defined in Section \ref{sec:model} and basic functionalities of the \scheme defined in Figure \ref{fig:functions}, we can define different security experiments as depicted in Figure \ref{fig:SecurityExp}. 
The security of the \scheme stems from the following assumptions: We assume that all deployed hash functions are secure according to definition \ref{def:hash}, all digital signatures are \eufcma-secure according to definition \ref{def:eufcma}, and the deployed dynamic accumulator is correct and sound according to definition \ref{def:accsound}. 
The security of enrollment, revocation, and update procedures stems from the PBFT consensus mechanism and the deployed accumulator. Hence, we can discuss how many honest nodes would be required to guarantee security. 

Let \thr denote the number of nodes in the consensus group where up to $f$ players are corrupted, and $h = t-f$ denotes the minimal number of honest players.  
For unconditionally secure protocols in the standard model with a complete and synchronous network of bilateral authenticated communication channels among the players and no trusted entity, a resilient broadcast is achievable if and only if $2\thr /3 < h$ \cite{ConsidineFFLMM05}. 
PBFT withstands against at most $f = \lfloor(\thr-1)/3\rfloor$ Byzantine nodes. Correctness proofs can be found in \cite{CastroL99c}. 
We assume that the network is controlled by an adversary that can reorder or delay messages under synchronous compliance and thus the protocol does not deviate from the synchrony assumption. 

\begin{customthm}{1} 
If the deployed hash function is collision-resistant, the deployed signature schemes are \eufcma-secure, the deployed accumulator is sound, and not more than $f=\lfloor(\thr-1)/3\rfloor$ nodes in the consensus group are Byzantine, then an adversary \advA has a negligible advantage in winning the security experiments described in Figure \ref{fig:SecurityExp}. 
\end{customthm}

\begin{figure}[!t]
\begin{center}
    $\underline{\Exp^{\atka}(\advA)}$ \\
    $(\id, \pk) \getsr \advA()$ \\ 
    $\ret 1 ~\iif (\register(\id, \pk , \flag, \role)=1) \wedge $\\ 
    $\hspace*{49pt}(\verify(\id, \pk) = \bot) \wedge (\pkveri(\id, \pk)=1) $ \\
\vspace{14pt}
    $\underline{\Exp^{\atkb}(\advA)}$ \\
    $(\id, \pk) \getsr \advA()$ \\ 
    $\ret 1 ~\iif (\register(\id, \pk , \flag, \role)=1) \wedge$ \\
    $\hspace*{49pt}(\verify(\id, \pk)= \bot) \wedge (\pkveri(\id, \pk) = \bot) $ \\
\vspace{14pt}
    $\underline{\Exp^{\atkc}(\advA)}$ \\
    $(\id, \pk) \getsr \advA()$ \\ 
    $\ret 1 ~\iif (\register(\id, \pk , \flag, \role)=1) \wedge $ \\
    $\hspace*{49pt}(\verify(\id, \pk)=1) \wedge (\pkveri(\id, \pk) = \bot) $ \\
\vspace{14pt}
    $\underline{\Exp^{\atkd}(\advA)}$ \\
    $(\id, \pk^*) \getsr \advA()$ \\ 
    $\ret 1 ~\iif (\update(\id, \pk^{old}, \pk^*)=1) ~\wedge $\\
    $\hspace*{39pt}(\verify(\id, \pk^*)=1) $ \\
\vspace{14pt}
    $\underline{\Exp^{\atke}(\advA)}$ \\
    $(\id, \pk) \getsr \advA()$ \\ 
    $\ret 1 ~\iif (\revoke(\id, \pk)=1) \wedge $\\ 
    $\hspace*{39pt}(\verify(\id, \pk) = \bot) $ \\
\end{center}
    \caption{Security definitions for the \scheme.} 
	\label{fig:SecurityExp}
\end{figure}

The theorem can be proved using sequences of games. In the first experiment $\Exp^{\atka}$, an adversary \advA wins if \advA can register a valid public key for an illegitimate entity. In the second experiment $\Exp^{\atkb}$, \advA wins if \advA can register an invalid public key for an illegitimate entity. \advA wins the third experiment $\Exp^{\atkc}$ if \advA can register an invalid public key for a legitimate entity. In the fourth experiment $\Exp^{\atkd}$, \advA wins if \advA can update/change the public key of a legitimate and enrolled entity with another public key of its choice. \advA wins the fifth experiment $\Exp^{\atke}$ if \advA can revoke the public key of a legitimate and enrolled entity without its consent. 
Note that the security of the first three experiments that concern the enrollment is equivalent, i.e. if \advA wins one of them, \advA will win the two others. 

\section{Conclusion}
A decentralized blockchain-based PKI (\scheme) was introduced in this paper that eliminates the single-point-of-failure and demanding needs for maintaining CA and revocation lists in the traditional PKIs. It distributes trust between entities through WoT and new keys are enrolled or revoked based on a consensus mechanism between trusted nodes that are already part of the system. Although we used PBFT, it can be replaced with more scalable and efficient variants. For efficient verification of public keys, a dynamic cryptographic accumulator is incorporated into the scheme that makes it suitable for IoT applications. \scheme provides different functionalities for registration, update, verification, and revocation of keys. All operations are transparent and auditable. 

\section*{Acknowledgment}
This research was supported by the Swedish Foundation for Strategic Research under Grant No. RIT-0035. 

\bibliographystyle{IEEEtran}
\small{\bibliography{PKI}}

\end{document}